# Nanostructured Multilayer Coatings for Spatial Filtering

Lina Grineviciute[1*], Ceren Babayigit[2], Darius Gailevičius[3,4], Martynas Peckus[3], Mirbek Turduev[5], Tomas Tolenis[1], Mikas Vengris[3], Hamza Kurt[2], Kestutis Staliunas[3,6,7*]

[1]Center for Physical Sciences and Technology, Savanoriu Ave. 231, LT-02300 Vilnius, Lithuania
[2]TOBB University of Economics and Technology, Söğütözü Str. 43, 06510, Ankara, Turkey
[3]Vilnius University, Faculty of Physics, Laser Research Center, Sauletekio Ave. 10, Vilnius, Lithuania
[4]Femtika, Sauletekio Ave. 15, LT-10224, Vilnius, Lithuania
[5]TED University, Ziya Gökalp Str. 48, 06420, Ankara, Turkey
[6]ICREA, Passeig Lluís Companys 23, 08010, Barcelona, Spain
[7]UPC, Dep. de Fisica, Rambla Sant Nebridi 22, 08222, Terrassa (Barcelona) Spain

*Corresponding authors: lina.grineviciute@ftmc.lt and kestutis.staliunas@icrea.cat

**Abstract**

Spatial filtering is an important mechanism to improve the spatial quality of laser beams. Typically, a confocal arrangement of lenses with a diaphragm in the focal plane is used for intracavity spatial filtering. Such conventional filtering requires access to the far-field domain. In microlasers, however, conventional filtering is impossible due to the lack of space in micro-resonators to access the far-field. Therefore, a novel concept for more compact and efficient spatial filtering is necessary.

In this study, we propose and demonstrate a conceptually novel mechanism of spatial filtering in the near-field domain, by a nanostructured multilayer coating - a 2D photonic crystal structure with a periodic index modulation along the longitudinal and transverse direction to the beam propagation. The structure is built on a nano-modulated substrate, to provide the transverse periodicity. The physical vapor deposition is used to provide self-repeating modulation in the longitudinal direction. We experimentally demonstrate a 5 μm thick photonic multilayer structure composed of nanostructured multiple layers of alternating high- and low-index materials providing spatial filtering in the near-infrared frequencies with 2° low angle passband. The proposed photonic structure can be considered as an ideal component for intracavity spatial filtering in microlasers.

**Introduction.** Spatial filtering of light beams is an essential procedure in laser science and technology. Cleaning of the spatial distortions of the laser beams can be performed in post-processing of the laser radiation to ensure the tight focusing or precise/controlled collimation of the light. However, even more important is the intracavity beam filtering inside of the laser resonators which enables the emission of maximally "clean" beam directly from the laser. In many kinds of lasers, the spatial distortions of the radiation are cleaned naturally due to the low aspect ratio of the system, when the longitudinal extent of the resonator is substantially larger than its transverse size, or due to resonators' curved mirrors with naturally occurring apertures. On the other hand, additional spatial filtering is sometimes performed by the intracavity spatial filter – a confocal lens arrangement with the aperture in the focal plane, which blocks higher-angle wave components (i.e., higher-order transverse modes) of the intracavity radiation. However, still for a large class of microlasers, such as semiconductor edge-emitting lasers, vertical-cavity surface-emitting lasers (VCSEL), microchip lasers, the intracavity spatial filtering is an unresolved issue. The natural beam cleaning mechanism is absent due to large aspect ratio of such micro-resonators, and the artificial filtering arrangement is impossible due to extremely small resonator dimensions (cavity length is in the order of millimeters, or in case of VCSEL – micrometers). A search for novel, efficient and convenient spatial filtering mechanisms, working in the near-field domain is crucial for the further development of such microlaser systems.

The spatial filtering in principle can be provided by photonic crystals (PhCs) - the structures with a periodically modulated refractive index on a wavelength scale, both longitudinally and transversely to the beam propagation direction[1-4]. Here, the filtering effect is based on the photonic bandgaps (BGs) of periodic media. Well known are the BGs in the frequency domain, which are due to the Bragg diffraction in layered (e.g., 1D index modulated) media, where the first BG occurs for the periods of modulation at around half-wavelength $\lambda/2$. Less evident is that also the angular BGs can occur in double-periodic (e.g., 2D index modulated) media, where the refraction index is modulated simultaneously in the longitudinal and transverse directions[5].

The principal possibility of spatial filtering by PhCs already has been discussed[1-4], but the experimental realization of the efficient spatial filters is still an unresolved issue. Volume refractive index modulation with the periods of around the $\lambda/2$ is technologically a great challenge. We note that the modulation must occur in 2D for 1D (cylindrical) spatial filtering, and in 3D for full (axisymmetric) spatial filtering. As the beams of several millimeter widths are to be cleaned in typical applications, the modulation must be perfectly uniform across the photonic structure over the ranges of millimeters. The transverse/longitudinal periods of the modulation must also be perfectly matched to provide the filtering on a given wavelength. Combining all these ingredients into one compact optical element is an unresolved challenge. Some attempts to fabricate spatial filters were made by direct laser writing in glasses[6,7], and although they even were successfully integrated into microlasers[8,9], they did not lead to spectacular improvement of the spatial structures of the radiation (the brightness increased 3 times in microchip lasers[8], and 1.5 times in edge-emitting lasers[9]).

There are also alternative ideas on compact spatial filtering in metasurfaces which can also provide angular sensitivity for plane waves reflecting/transmitting through the structure[10]. Various angular dispersion properties were demonstrated giving rise to incident-angle-selective meta-absorber, an angle-multiplexed meta-polarizer, an angle-multiplexed wavefront controller, and angle-multiplexed broadband metasurface sensor[11,12]. The aforementioned works are promising in perspective; however, they are still far from the technological applications in real photonic systems and high-power microlasers.

In this study, we propose and realize the spatial filtering effect based on the mechanism of the angular BGs in double-periodic photonic structures (see the schematic of the structure in Fig. 1). Practically, the transverse periodicity results in a translation-invariant behavior, therefore such filters should be easy to align considering the fact that aligning optics precisely can be a very challenging task. From the technological viewpoint, the work combines lithographical methods to provide the seed of transverse modulation and the physical vapour deposition approach for longitudinal refractive index modulation. The transverse modulation was obtained by structuring the surface of the substrate by laser interference lithography, and the ion-beam sputtering process was used to conformally deposit multilayer coating on pre-microstructured substrates. There were also previous attempts to coat the micro-structured substrates[5,13], but no implementation has been presented. The thin films replicate the transverse seed modulation of the substrate and, as a result, the double-periodical modulation of the refractive index is obtained, which is a necessary ingredient for the PhC spatial filtering. The present article elaborates on the physical idea of filtering with its technological idea of fabrication. Moreover, the study presents a systematic search of the optimum geometry by simulations in finite-difference time-domain (FDTD) and plane-wave expansion (PWE) methods, an analytical approach of the design, fabrication using lithographic methods, multilayer coatings, and experimental measurements of spatial-angular transmission of the design. One of the main messages of the article is the first experimental demonstration of the functional 2D PhC spatial filter of several micrometer thicknesses for the near-infrared radiation.

**Design.** The geometry of the photonic structure is schematically shown in Fig. 1. The structure consists of alternating thin films with high and low refractive indices on a modulated substrate. The curved thin films replicate the surface modulation of the substrate, providing the refractive index modulation in both, the longitudinal (z-axis) and transverse (x-axis) directions. Note that the refraction from such a structure can be classified as the generalized Bragg one since the refracted components propagate in a backward direction. The condition for the back-refraction is that the longitudinal component of the index modulation should have a periodicity of $\lambda/2 < d_z < \lambda$. This is enabled in very compact dimensions of the filtering structure. For instance, 15 periods (30 layers) of modulation result in layered coatings of approximately $10\,\lambda$ thickness, which for the near IR, considering the refractive index of the material, is around $5\,\mu m$. This is a significant advantage of our filtering principle compared with the spatial filters in Laue configuration[3,4,8,9], where the thickness of the spatial filter is of the order of 1 mm. Moreover, in Laue forward refraction configuration no "true" BGs are possible – just the quasi-BGs appear, as the refraction is directed predominantly in the forward direction. Furthermore, this thin arrangement of the filtering layers can support resonant wave-guiding modes if the coating has a higher refractive index than its surrounding media. Depending on the incidence angle and the wavelength of the incoming light, waveguiding modes can lead to enhanced/reduced transmission as a result of constructive/destructive interference, and may improve the BG-based filtering effect.

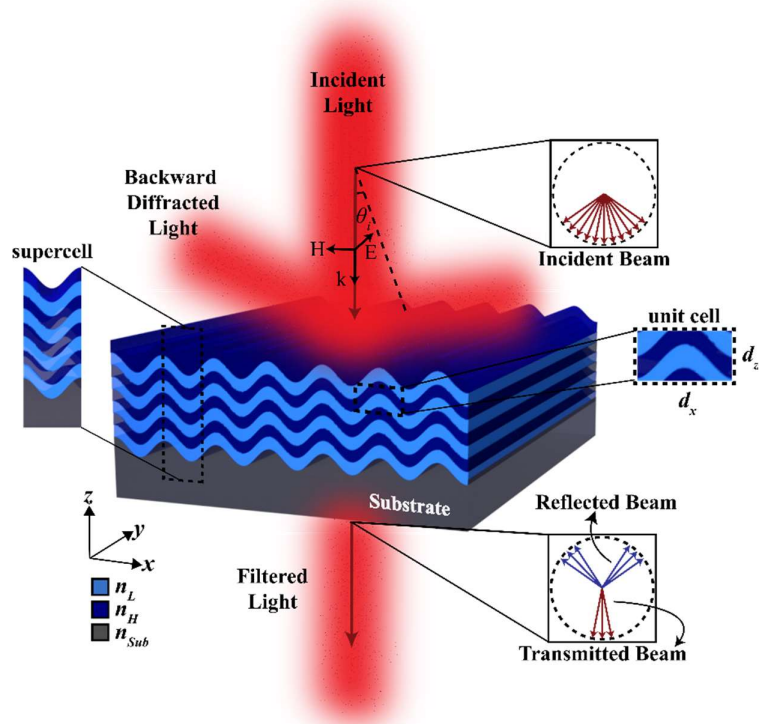

*Fig. 1 Photonic structure consisting of a self-replicating curved multilayer coating on a periodically modulated substrate. The insets show the elementary cell to calculate the photonic band diagrams, and the supercell used to calculate (angle, wavelength) diagrams of field transmission through the structure. The insets also illustrate the double-refraction based angle-selectivity of the transmission, providing the spatial filtering. Electric field is oriented for s-polarization case.*

The idea of the angular BGs filtering is first explored by the calculations of the band diagrams of the proposed structure by using the standard PWE method. To extract the band diagram of the structure, we follow the ΓY trajectory (*i.e.*, along the beam propagation direction), then YM, and then back MΓ in the reciprocal-space, see Fig. 2(a), for the *s*-polarized electromagnetic field where the electric field is directed parallel to the groves. Here, the BGs along the optical axis (ΓY-direction) is important for filtering effect (see the inset dashed windows in the band-diagram, Fig. 2(a)). In the middle of the axial BG, evidently, the axial components are blocked, which hints on the high-angle-pass spatial filtering effect. However, for the frequencies slightly below and above the axial BG the low-angle-pass filtering can be expected.

The scenario has been explored calculating the iso-frequency contours (IFCs) for the frequencies around the axial BG. The IFCs in Figs. 2(b)-2(c) are combined from the 2$^{nd}$, 3$^{rd}$, and 4$^{th}$ photonic bands, since the same frequency IFCs can be found in different bands. Indeed, for the frequencies inside the axial BG, the high-angle-pass filtering is obtained, see Fig. 2(b), and for the frequencies slightly below and above the BG low-angle-pass filtering occurs as can be seen in Figs. 2(c) and 2(d), respectively. A complete collection of relevant IFCs is presented in Suppl. 1. Interestingly, the same scenario of low-high-low angle-pass filtering occurs also around the second frequency BG, although to a lesser extent, as indicated in Fig. 2(a) and documented in Suppl. 1.

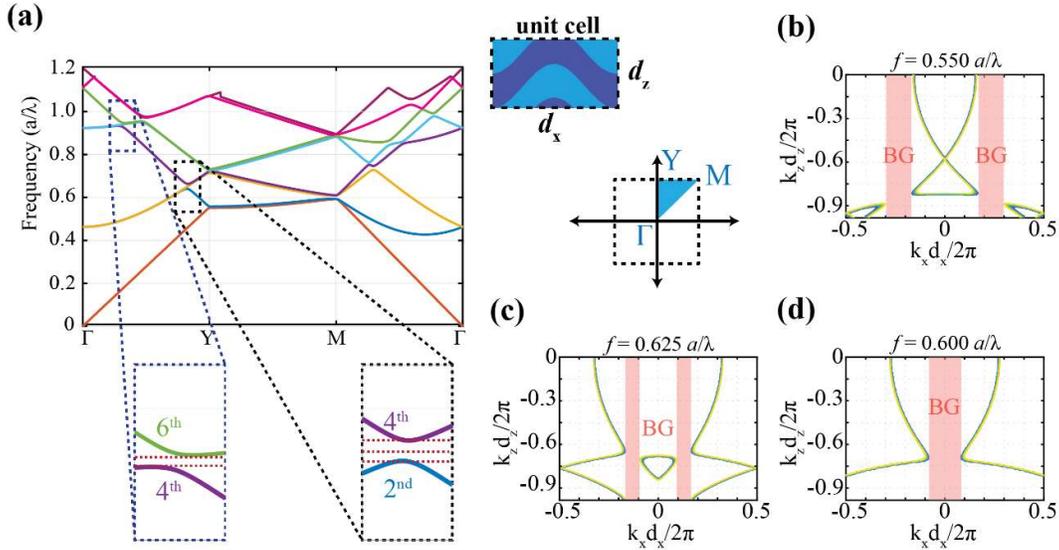

*Fig. 2 a) Band diagram of the (infinitely extended) periodic structure, with the regions of interest zoomed out (only 2$^{nd}$, and 4$^{th}$ bands indicated in the first, and 4$^{th}$, and 6$^{th}$ bands in the second inset window). IFCs combined from 2$^{nd}$, 3$^{rd}$, and 4$^{th}$ bands, indicating high-angle-pass filtering at the middle of the first frequency BG (b), and low-angle-pass filtering below and above the first BG (c, d), respectively. Parameters: transverse period $d_x$=600 nm, longitudinal period $d_z$=301 nm, modulation depth 150 nm, refractive indices $n_L$=1.99 and $n_H$=2.24.*

Next, we explore numerically the transmission of the plane waves through the photonic structure depending on their incidence angle and the wavelength. Due to the transverse periodicity, we used the supercell analysis of the field propagation by the FDTD method, with Bloch-periodic boundary condition on the lateral boundaries of the supercell[14,15]. Corresponding results are summarized in Fig. 3(a). Here, we present the results only for the *s*-polarized incident wave (electric field orientation is parallel to the fringe direction, *i.e.* parallel to the *y*-axis). The *p*-polarization gives qualitatively the same response, see Suppl. 2. Furthermore, analytically calculated angular-frequency bandgaps are presented and superimposed in Fig. 3(b) (for detailed analytical derivations see Suppl. 3). Also, the cross-sectional profiles of the transmittance maps are extracted for wavelengths of $\lambda$ = 1051 nm and $\lambda$ = 972 nm in Figs. 3(c) and 3(d), respectively. Additionally, a vertical cross-sectional profile indicating the frequency BGs at the incidence angle $\theta_i=0°$ is presented in Fig. 3(e). As can be seen from Figs. 3(c) and 3(d), almost perfect transmission occurs within small incidence angles and the waves at slightly larger angles are strongly suppressed in the transmitted light. Interestingly the second frequency BG also provides spatial filtering but with comparably lower efficiency, at around the wavelength of $\lambda = 620$ nm, however, a detailed study was not performed.

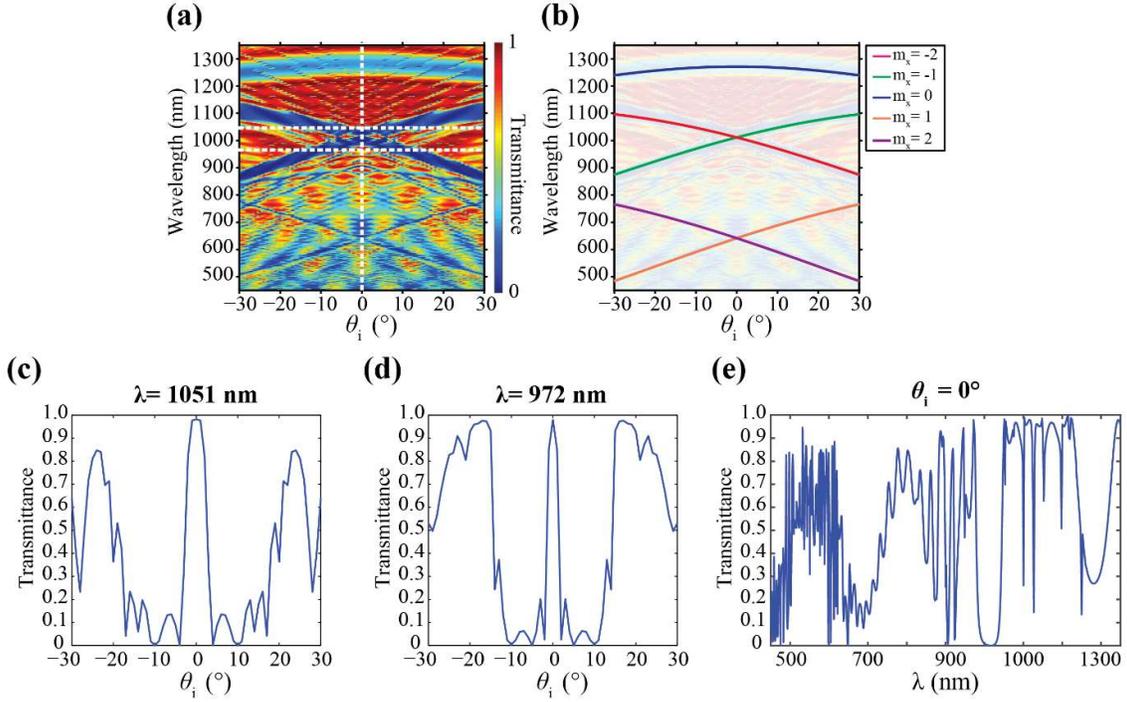

*Fig. 3 a) Map of transmission dependence on incidence angle and the wavelength extracted from the FDTD numerical calculations on a supercell. Two "crosses" of Bragg reflection lines are visible indicating the first and the second axial BG. Two dashed lines indicate cross-sectional transmission profiles of $\lambda = 1051$ nm and $\lambda = 972$ nm. (b) Analytically calculated Bragg bandgap positions. Angular dependences of transmissions at different wavelengths of (a) $\lambda=1051$nm and (b) $\lambda=972$nm (horizontal cross-sections of the map) showing efficient spatial filtering at around the first filtering BG. (e) Vertical cross-section at the incidence angle $\theta_i =0°$ at the transmission map in (a). Here, the structural parameters are the same as in Fig. 2 and the number of layers equal to N=33.*

The overall filtering pattern (*i.e.*, angular BGs positions presented in Fig. 3(a)) corresponds well with the analytically calculated ones (Fig. 3 (b)), see the Suppl. 3. However, additionally, the transmission spectrum in the map was overlaid by the fringed structure. Vertical fringing is evidently related to the Fabry-Perot resonances of the coating structure, as the frequency separation between the fringes corresponds to the resonances of the structure of 5 μm geometrical thickness (10 μm optical thickness). The origin of the horizontal part of fringing is more sophisticated. We identify them as the resonances of the first order diffracted modes trapped in the planar structure of a relatively high refraction index. The corresponding waveguided modes are calculated in Suppl. 4, resulting in a correct angular period of fringes. In addition to the main mechanism of the angle-selective refraction effect on a grating, the wave-guiding effect can increase/decrease spatial filtering depending on the parameters (essentially on the thickness parameter of the coating). Further optimization of this additional effect can be used to improve the filtering performance of this type of spatial filter.

**Fabrication and characterization.** The substrates with periodically-transversally modulated surfaces were fabricated by laser interference lithography and nano-imprint technology. A chain of technological processes was applied to fabricate the samples. The fabrication process consists of the following steps: *i)* preparation of the master copy using the interference method in a photoresist *ii)* imprinting of the master structure on the substrate (using the UV imprint polymer); *iii)* physical vapor deposition on the modulated substrate to form the multilayered structure. The details are provided in the methods section as well as in Suppl. 5.

The inner structure of the sample was characterized by using Scanning Electron Microscope (SEM). In order to analyze the cross-sections, samples were cut by a laser beam and broken in the middle of the sample. Additionally, surface morphology was measured by Atomic Force Microscope (AFM) before and after depositions of the coating to precisely determine the depths of the curved structure modulation, see Fig. 4(a).

Transmission maps for the fabricated sample were recorded by spectrophotometric measurements. Here, linearly polarized light was used for two perpendicular polarizations: S and P, where S polarization is parallel to the grating lines on the sample. The angle between the main plane of the grating and the detector was varied from -30° to 30° by steps of 1°. The resulting transmission map is presented in Fig. 4(b).

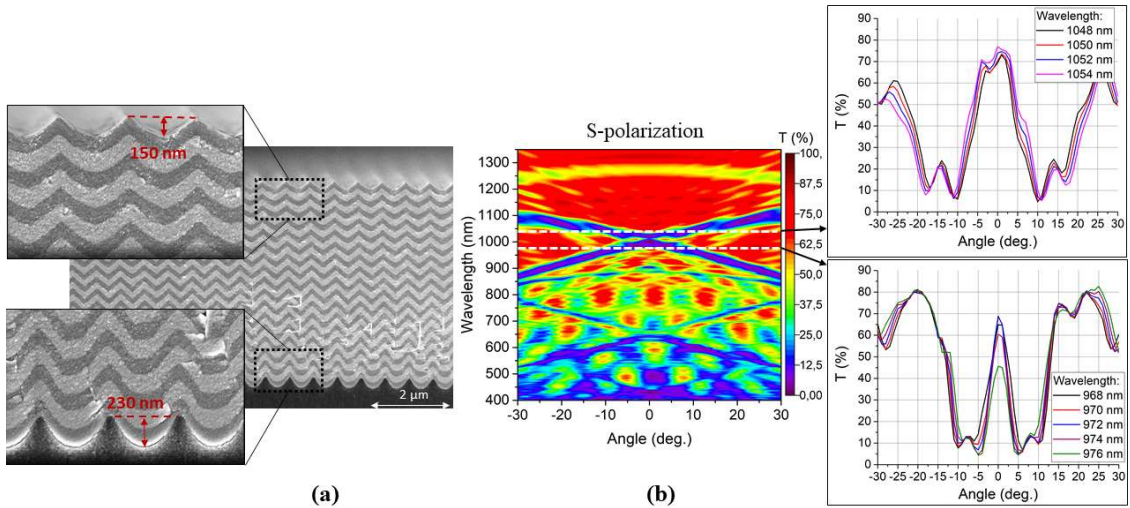

*Fig. 4 a) SEM images of the cross-section of the structure. b) Transmission map (transmission depending on the angle (horizontal axis) and wavelength (vertical axis)), together with cross-sections of the map (angular transmission curve) for characteristic wavelengths in the insets.*

The transmission maps correspond quantitatively to the ones obtained by numerical FDTD simulations. The main feature of the map – the crossings of the Bragg-reflection lines, is evident. Also, the fringing is visible in the measured map. Similarly, to the numerical results, two crossings of the Bragg-reflection lines were observed, the main cross indicating the strong spatial filtering associated with the first BG and the weaker one associated with the second BG.

It should also be noted that SEM images of the cross-sections indicate a general tendency where the curving amplitude of the layers decreases with an increasing number of layers (see the inset cross-sections in Fig. 4(a)). Typically, the modulation of the substrate with the amplitude of ~220 nm decreases to the amplitude in the range of 110 nm – 150 nm at the top layer surface. The influence of the sweep of the modulation amplitude was also examined with the FDTD supercell analysis, and the obtained outcome showed that it did not essentially affect the results. The FDTD simulations with the modulation sweep lead to a weak deformation of the transmission maps without leading to essential modifications. The comparison is provided in Suppl. 6.

The main quantitative difference between numerical results and spectroscopic measurements of the fabricated structure is that the filtering lines in the latter case were less deep and more blurred. This can be attributed to the imperfections of fabricated structure – the corrugation of the substrate modulation on a range of around 10 µm and, subsequently, the corrugation of the whole structure. The influence of these large-scale imperfections is estimated in Suppl. 7.

**Direct proof of Spatial Filtering.** Finally, the fabricated sample was illuminated by a cylindrically focused laser beam of a tunable picosecond laser (details of the laser device are provided in the Methods section). Figures 5(a) and 5(b) compare the far field (angular) profiles of the incident probe beam and the resulting filtered beam for slightly different wavelengths. The narrowing of the angular spectra in the transmission is evident – angular divergence of the beam at FWHM is ~ 2°. This is the main result of the article – specifically direct experimental proof for the spatial filtering effect by nanostructured multilayer coating.

For a more detailed analysis, a transmission map was reconstructed by dividing the measured angular intensity profile of the filtered beam by the reference (unfiltered) one. For comparison, we show the reconstructed transmission map in Fig. 5(c) together with the one that obtained from spectrophotometric data in Fig. 5(d). Angular transmission spectra for two wavelengths of $\lambda$ = 968 nm and $\lambda$ = 972 nm are shown in Figs. 5(e) and 5(f) for an additional comparison. The first case shows a transmission curve for a more bell-shaped profile with a higher transmission loss at the low angle values. On the other hand, the second case shows a lower loss case with a flat top resembling case. The detailed record of measurement data is provided in Suppl. 8.

**Discussion.** The main message of the article is the demonstration of pronounced functional spatial filtering by thin nanostructured multilayer coating and with detailed numerical, analytical and experimental data, we are able to demonstrate such filtering of light. However, some imperfections still prevent the filter from technical applications, which should be discussed below and could be tackled in future work.

One issue is the broadening of the filtering lines due to the technical artifacts – the corrugation of the substrate on a large scale, and of the structure itself. With the current technology of the substrate fabrication, this limits the half-width of the filtering line approximately to 2°. As numerical simulations show, the removal of corrugation would allow reducing the half-width of filtering lines to approximately 1° for the current configuration.

The other issue is that the demonstrated filtering is a cylindrical one. The extension to 2D is, however, straightforward and consists in a change of the substrate modulation geometry (to square, hexagonal, concentric, etc.). The filtering geometry depends solely on the substrate, as the subsequent coating of layers in the vertical direction is independent of the geometry of the modulation of the substrate.

Hence, the fabricated multilayer structure (1D filtering, 2 degrees half-angle) could be well applied for broad area semiconductor edge emission lasers, where the reduction of the beam divergence from typical 5°-10°, to desired 1°-2° would be very profitable. The filters for the other relevant systems can require different angular width of the transmission. To reduce the width (to obtain narrower angular transmission lines) generally the index contrast of deposited layers should be decreased, and the number of layers is increased.

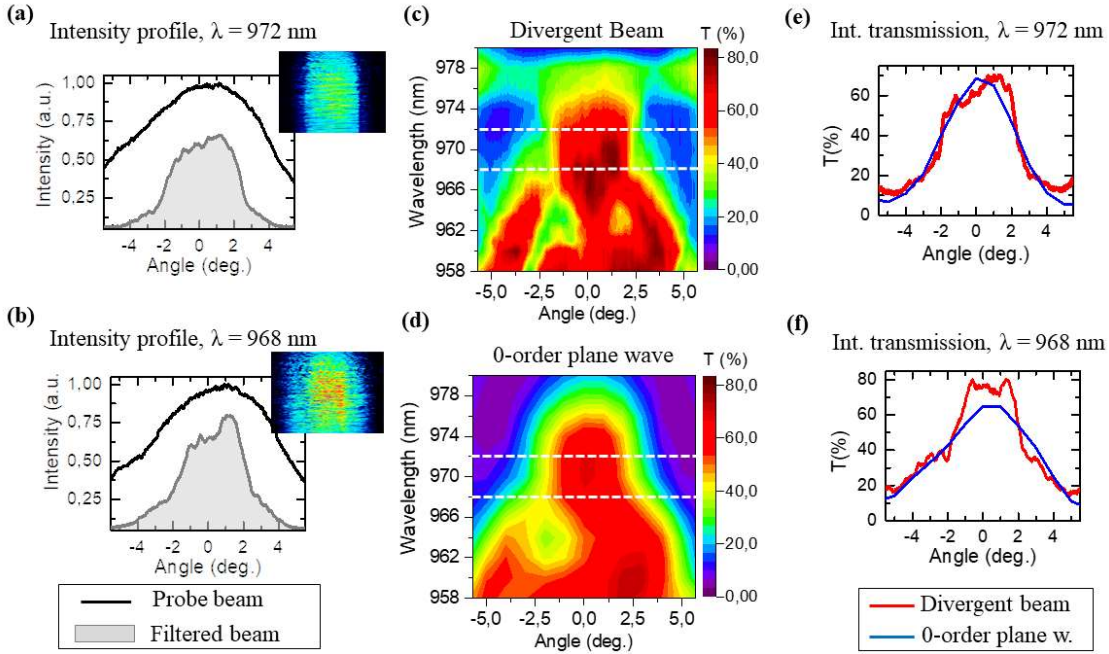

*Fig. 5 Experimental data for spatial filtering of a divergent probe beam. (a, b) Transmission of a reference beam and filtered beam evidencing 1D (cylindrical) spatial filtering for two different wavelengths. Insets show the 2D intensity profiles of the filtered beam, registered on a CMOS camera. (c) Shows the corresponding transmission map, where the sample was illuminated with a divergent probe beam. Wavelengths of interest are marked with dotted lines. (d) Shows the zoomed-in range from Fig. 4(b) corresponding to a 0-order plane wave transmission spectral-angular map. (e) and (f) show the intensity transmission profiles for wavelengths 972 and 968 nm for both measurement methods.*

**Methods and Materials.** A variety of methods were used to achieve the main result – the demonstration of spatial filtering.

1) **Design** of the structure is based on numerical simulations using standard methods: plane wave expansion in a perfectly periodic structure to calculate the frequency isolines; FDTD simulations on a supercell to calculate the (angle, wavelength) transmission maps. Both are well established methods[15,16]. For the proof of the spatial filtering, the full FDTD simulations on a complete structure were performed, which is also a well-established method.

2) **Fabrication**. The laser interference lithography, employing the third harmonics of Nd:YAG laser radiation, was used to form the master grating. Subsequently, the master was used in a typical soft nano-imprint lithography process[17,18], where an intermediate soft polydimethylsiloxane imprint was taken from it and used for UV cured polymer (OrmoComp®, $n_{ref}$ = 1.52) copies production. The resulting samples with sinusoidally modulated surfaces (600 nm transverse period and 220 nm modulation depth) were used as the substrate for multilayer coating. The coating consists of alternating high and low refractive index thin films, deposited by the Ion Beam Sputtering method. Specifically, hafnia and niobia materials were used with refractive indices $n_L$ = 1.99 and high $n_H$ = 2.24, respectively. The multilayer structure consists of 33 layers, whose optical thicknesses were equal for both the high and low refractive index layers – 317 nm ($d_z$ = 301 nm with the geometric thicknesses – 141.8 nm and 159.2 nm, respectively). More technical details are provided in Suppl. 5.

3) **Characterization** was performed by the AFM analysis of the surface's morphology of the substrate and of the upper layer of the fabricated structure; SEM was used to explore the cross-section of the fabricated structure; Finally, the spectroscopical measurements of the transmission was performed in order to build the (angle, wavelength) transmission maps.

4) **Demonstration** of spatial filtering by recording the far field profiles of the transmitted beam from a tunable picosecond laser source, tuned in the wavelength range of 958 nm ≤ λ ≤ 980 nm using a monochromator with a spectral bandwidth of approximately $\Delta\lambda \approx$ 1 nm. Focusing provides the angular divergence of the beam of ~ 9.7° at FWHM (full width at half maximum) in the horizontal direction. The resulting filtered beam was registered with a CMOS beam profiler camera while imaging the beam directly onto the sensor.

**Acknowledgments:** This project has received funding from European Social Fund (project No 09.3.3- LMT-K712-17- 0016) under grant agreement with the Research Council of Lithuania (LMTLT), D.G. and M.P. acknowledge the financial support from "FOKRILAS" (Project No. P-MIP-17-190) from Research Council of Lithuania. Authors are grateful to dr. Algirdas Selskis from State research institute Center for Physical Sciences and Technology for SEM measurements.

**Author contributions:** K.S. supervised the project and provided theoretical analysis. C.B., M.T., H.K. performed the simulations. L.G., D.G. fabricated modulated substrates, L.G., T.T., D.G. designed and carried out the experiments and measurements. M.P., M.V. performed measurements with a divergent laser beam. All authors participated in discussions and reviewed the manuscript.

**Data availability:** All data files that support this research are available from the corresponding author upon request.

**Conflict of interest:** The authors declare that they have no conflict of interest.

## References

1. K. Staliunas and V. J. Sánchez-Morcillo, "Spatial filtering of light by chirped photonic crystals," Phys. Rev. A 79(5), 053807 (2009).
2. E. Colak, A. O. Cakmak, A. E. Serebryannikov, and E. Ozbay, "Spatial filtering using dielectric photonic crystals at beam-type excitation," J. Appl. Phys. 108(11), 113106 (2010).
3. L. Maigyte, T. Gertus, M. Peckus, J. Trull, C. Cojocaru, V. Sirutkaitis, and K. Staliunas, "Signatures of light-beam spatial filtering in a three-dimensional photonic crystal," Phys. Rev. A 82(4), 043819 (2010).
4. V. Purlys, L. Maigyte, D. Gailevičius, M. Peckus, M. Malinauskas, and K. Staliunas, "Spatial filtering by chirped photonic crystals," Phys. Rev. A 87(3), 033805 (2013).
5. L. Grinevičiūtė, C. Babayigit, D. Gailevičius, E. Bor, M. Turduev, V. Purlys, T. Tolenis, H. Kurt, K. Staliunas, "Angular filtering by photonic microstructures fabricated by physical vapour deposition," Applied Surface Science, **481**, 353 (2019).
6. L. Maigyte and K. Staliunas, "Spatial filtering with Photonic Crystals," Applied Physics Reviews, **2**, 011102 (2015).

7. D. Gailevičius, V. Purlys, and K. Staliunas, "Photonic Crystal Spatial Filters Fabricated by Femtosecond Pulsed Bessel Beams, Optics Letters," **44**, 4969-4972 (2019).
8. D. Gailevicius, V. Koliadenko, V. Purlys, M. Peckus, V. Taranenko, and K. Staliunas, "Photonic Crystal Microchip Laser," Sci. Rep. 6(1), 34173 (2016).
9. S. Gawali, D. Gailevičius, G. Garre-Werner, V. Purlys, C. Cojocaru, J. Trull, J. Montiel-Ponsoda, and K. Staliunas, Photonic Crystal Spatial Filtering in Broad Aperture Diode laser, Appl. Physics Letters, **115**, 141104 (2019).
10. J. Cheng, S. Inampudi, and H. Mosallaei, "Optimization-based dielectric metasurfaces for angle-selective multifunctional beam deflection," Sci. Rep., 7, 12228 (2017).
11. N. Born, I. Al-Naib, C. Jansen, R. Singh, J. V. Moloney, M. Scheller, and M. Koch, "Terahertz Metamaterials with Ultrahigh Angular Sensitivity," Adv. Optical Mater. 3(5), 642-645, (2015).
12. A. Leitis, A. Tittl, M. Liu, B. H. Lee, M. B. Gu, Y. S. Kivshar and H. Altug, "Angle-multiplexed all-dielectric metasurfaces for broadband molecular fingerprint retrieval," Science Advances, 5(5), 1-8 (2019).
13. J. B. Oliver, T. J. Kessler, B. Charles, and C. Smith, Fabrication of a Continuous-Enfolded Grating by Ion-Beam–Sputter Deposition, presented at SVC Techcon 2014, Chicago, IL, 3–8 May 2014.
14. A. Taflove and S. C. Hagness, Computational Electrodynamics (2005).
15. "Lumerical FDTD Solutions, Inc., http://www.lumerical.com for high performance FDTD-method Maxwell solver for the design, analysis and optimization of nanophotonic devices, processes and materials.," (n.d.).
16. S. G. Johnson and J. Joannopoulos, "Block-iterative frequency-domain methods for Maxwell's equations in a planewave basis," Opt. Express, 8(3), 173-190 (2001).
17. T. Tamulevičius, S. Tamulevičius, M. Andrulevičius, E. Griškonis, L. Puodžiukynas, G. Janušas, A. Guobienė, "Formation of periodical microstructures using interference lithography," Experimental Techniques, 32, 23-28 (2008).
18. Y. Xia, G.M. Whitesides, "Soft lithography," Angew. Chemie Int. Ed. 37 (5), 550–575 (1998).

**List of supplementary material (below):**

**Supplement. 1:** Frequency isolines

**Supplement. 2:** Spatial filtering for *s* and *p* polarizations

**Supplement. 3:** Frequencies of the Bragg-filtering lines

**Supplement. 4:** Fringing in transmission maps

**Supplement. 5:** Sample preparation

**Supplement. 6:** Effects of modulation sweep

**Supplement. 7:** Effects of the corrugation of the substrates

**Supplement. 8:** Beam transmission data

# Supplements to:

# Nanostructured Multilayer Coatings for Spatial Filtering

Lina Grineviciute[1*], Ceren Babayigit[2], Darius Gailevičius[3,4], Martynas Peckus[3], Mirbek Turduev[5], Tomas Tolenis[1], Mikas Vengris[3], Hamza Kurt[2], Kestutis Staliunas[3,6,7]

[1]Center for Physical Sciences and Technology, Savanoriu Ave. 231, LT-02300 Vilnius, Lithuania
[2]TOBB University of Economics and Technology, Söğütözü Str. 43, 06510, Ankara, Turkey
[3]Vilnius University, Faculty of Physics, Laser Research Center, Sauletekio Ave. 10, Vilnius, Lithuania
[4]Femtika, Sauletekio Ave. 15, LT-10224, Vilnius, Lithuania
[5]TED University, Ziya Gökalp Str. 48, 06420, Ankara, Turkey
[6]ICREA, Passeig Lluís Companys 23, 08010, Barcelona, Spain
[7]UPC, Dep. de Fisica, Rambla Sant Nebridi 22, 08222, Terrassa (Barcelona) Spain

**Supplementary 1:** Frequency isolines

Here we present isofrequency lines of the spatial dispersion surface ($k_x$, $k_z$) for the investigated photonic structure. Isofrequency diagrams are calculated considering that the structure is perfectly periodic and infinitely extended. Fields are expanded into plane waves, and the Maxwell equations are solved on a unit cell (see Fig. 1(a) in the manuscript). The isofrequency diagrams were calculated both, for *s*-polarization (electric field orientation is parallel to the fringe direction, i.e. parallel to the *y*-axis) and *p*-polarization. The results are presented in Fig. S1 and Fig. S2, respectively. We plot the frequency isolines from 1st to 6th photonic band.

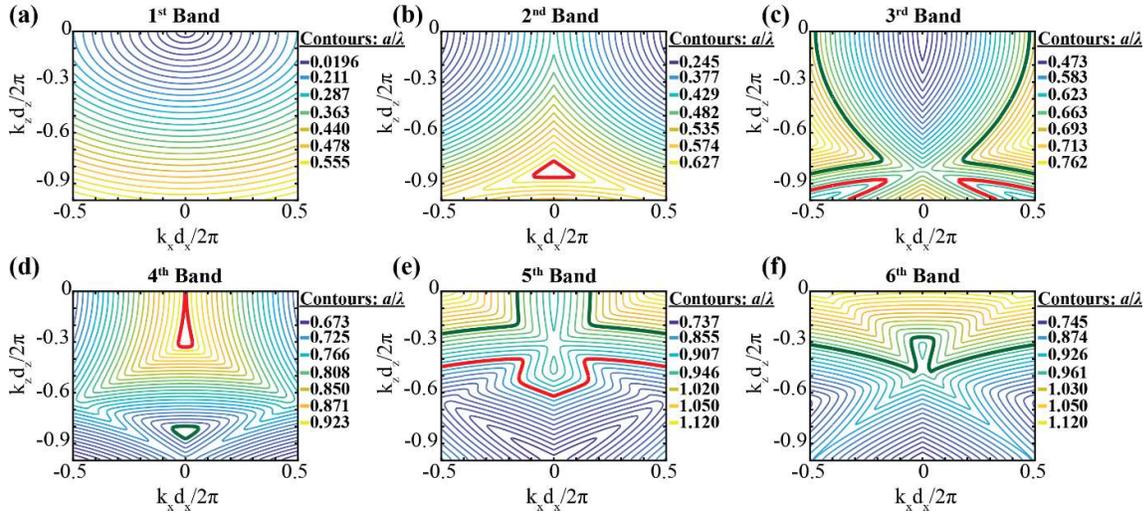

Figure S1. (a)-(f) Collection of isofrequency lines for different photonic bands, from 1st till 6th for *s*-polarization. The thick red lines from the 2nd and 3rd bands form the first low-angle-pass filtering area (just below the first frequency BG). Thick green lines from the 3rd and 4th bands form the second low-angle-pass filtering area (just above the first frequency BG). The red-green lines from 4th, 5th and 6th band indicate the low-angle-pass filtering areas around the second frequency BG. The parameters are: transverse period $d_x$=600 nm, longitudinal period $d_z$=301 nm, modulation depth 150 nm, the indices of refraction $n_L$=1.99 and $n_H$=2.24, which correspond to the structure from the main manuscript.

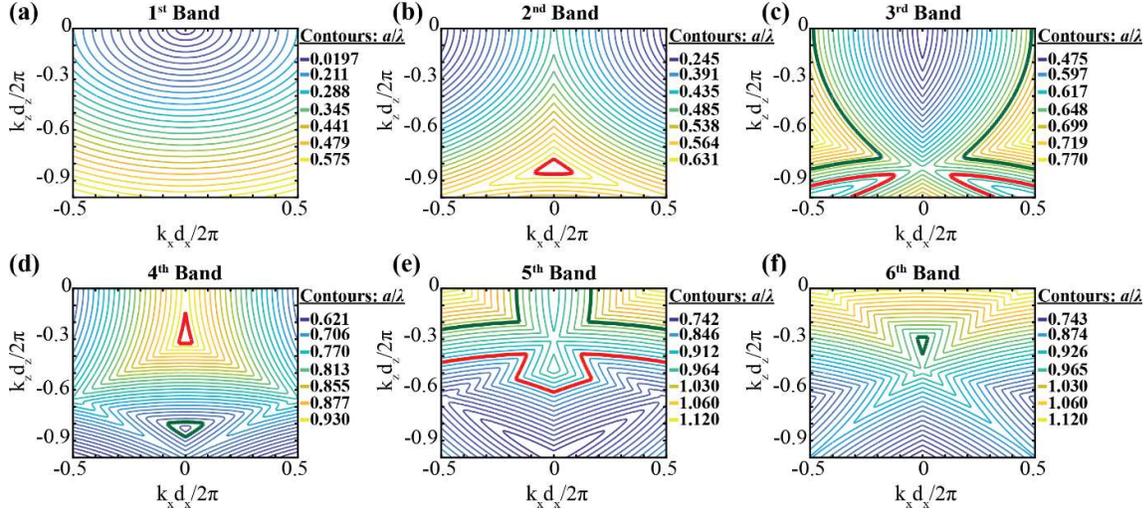

Figure S2. (a)-(f) Collection of isofrequency lines for different photonic bands, from the 1st till 6th for *p*-polarization. The parameters are as in Fig. S1.

The angular bandgaps can be identified by combining isofrequency lines from several photonic bands. For instance, the first relevant low-angle-pass angular bandgap (for the frequencies just below the first frequency bandgap) can be identified from the isolines in 2nd and 3rd bands (indicated by the thick red line). The second relevant low-angle-pass angular bandgap (for the frequencies just above the first frequency bandgap) can be identified from the isolines in 3rd and 4th bands (indicated by the thick green line). As can be seen from Fig. S1 and Fig. S2, the topology of isofrequency lines is the same for both s and p polarization, except for weak distortions and frequency shifts.

Further, the isofrequency lines in higher-order bands allow identifying the second spatial filtering area. For instance, isolines from 4th and 5th bands form the low-angle-pass angular bandgap for the frequencies just below the second frequency bandgap, and the isolines from 5th and 6th bands form the low-angle-pass angular bandgap for the frequencies just above the second frequency bandgap, as indicated by thick lines.

**Supplementary 2:** Spatial filtering for *s*- and *p*- polarizations

In this section, we present transmission maps with both *s*- and *p*- polarized incident beams for the structure considered in the main manuscript. Here, the numerical calculations of the structure were performed via the FDTD method on a supercell with the Bloch-periodic lateral boundary conditions (see the main text for the definition of supercell). The parameters are as in the main manuscript: the transverse period is fixed to 600 nm and the number of layers is $N_L$=33. *Hafnia* and *niobia* materials were used as low ($n_L$=1.99) and high ($n_H$=2.24) refractive index materials, respectively. The optical thicknesses of the high and low refractive index layers are both set to 317 nm ($d_z$=301 nm with the geometric thicknesses - 141.8 nm and 159.2 nm, respectively). The proposed structure is illuminated by both *s*- and *p*- polarized plane wave sources at adjustable incident beam angles of -30°<$\theta_i$<30° and the corresponding zero-order transmission is calculated in the range of the wavelengths between 450 nm and 1350 nm. The obtained results for both *s*- and *p*- polarizations are presented in Fig. S3, together with the results of experimental measurements. Numerical calculations and experimental measurements indicate that the geometry of the filtering Bragg lines is the same for both polarizations, they only show the quantitative differences (i.e., different "depth" of the Bragg lines).

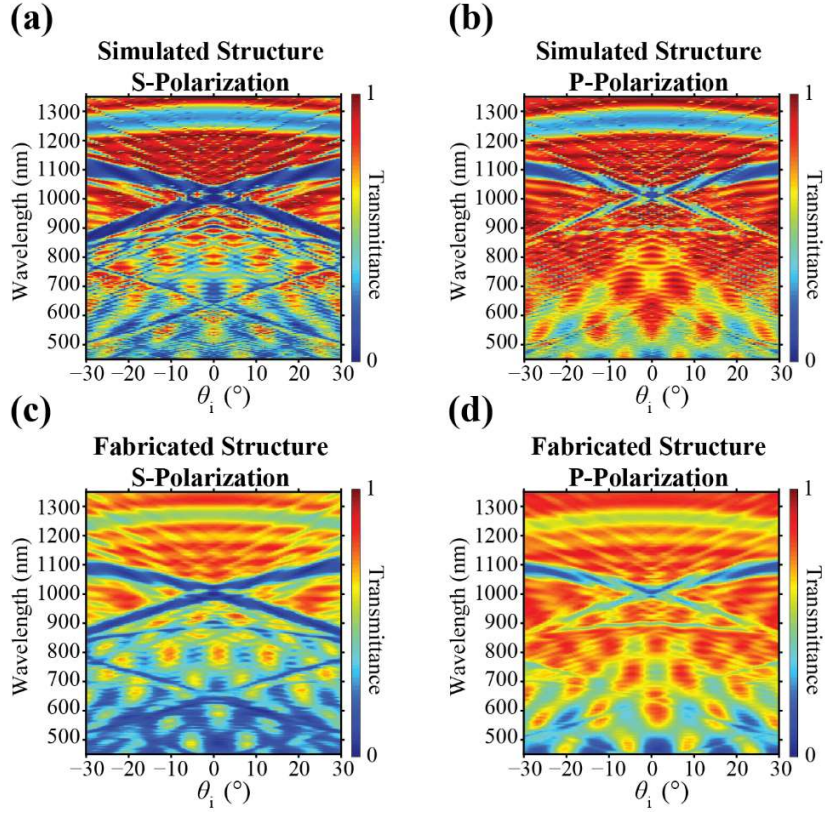

Figure S3. Zero-order transmittance maps calculated for *s*-polarization (a) and *p*-polarization (b). Experimentally measured zero-order transmittance maps for (c) *s*- and (d) *p*- polarizations.

**Supplementary 3:** Frequencies of the Bragg-filtering lines

The frequencies for the generalized Bragg bandgaps can be derived from the resonant scattering condition using a simple geometric approach. An incident wave propagating at the angle of $\theta_i$ to the normal of the substrate $\vec{k}_0 = (k_0 \sin(\theta_i), k_0 \cos(\theta_i))$, diffracting on a grating with $\vec{q} = (q_x, q_z)$, results in a diffracted wave $\vec{k} = \vec{k}_0 + \vec{q} = (k_0 \sin(\theta_i) + q_x, k_0 \cos(\theta_i) - q_z)$. Here $|\vec{k}_0| = \omega/c$ is the modulus of the wavevector, and the components of the lattice vector are $q_{x,z} = 2\pi/d_{x,z}$ (see Fig. S4. for the definition of the transverse/longitudinal periods $d_{x,z}$). For efficient diffraction, the new wave with wave-vector $\vec{k}$ must propagate, i.e., must satisfy $|\vec{k}| = \omega/c = |\vec{k}_0|$. This means that the initial, as well as final, wave-vectors must lie on the "light ring": $|\vec{k}|^2 = (\omega/c)^2$. This resonance condition $|\vec{k}| = |\vec{k}_0|$ leads to the following implicit condition for the Bragg diffraction angle:

$$|\vec{k}| = \frac{q_x^2 + q_z^2}{2q_z \cos(\theta_i) - 2q_x \sin(\theta_i)} \tag{S. 1}$$

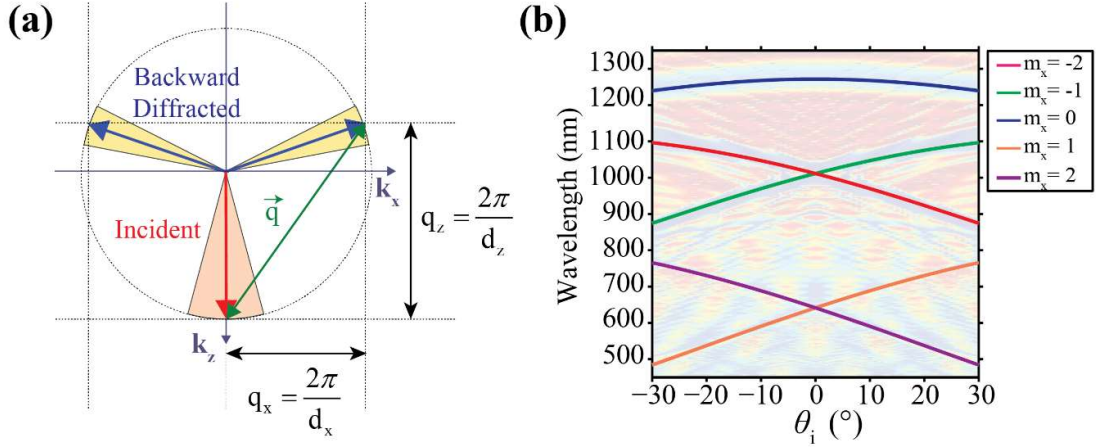

Figure S4. (a) Geometry of wavevectors in Bragg diffraction. (b) Bragg-wavelength dependences on incidence angle as follows from Eq (S.4) with the parameters $n$=2.11, $d_x$=600 nm, $d_z$=301 nm, with $m_x$=-2, -1, 0, 1, 2, and $m_z$=1.

Respectively the wavelength:

$$\lambda = 2\frac{\cos(\theta_i)/d_z - \sin(\theta_i)/d_x}{(1/d_x)^2 + (1/d_z)^2} \tag{S. 2}$$

The expression in Eq. S.2 is a convenient tool to interpret the angular diffraction map, i.e., the dependence of the diffracted wavelength on the incidence angle of $\theta_i$. It shows one Bragg diffraction curve in transmission map, i.e., the diffraction due to the main harmonics $\vec{q} = (q_x, q_z)$ of the periodic structure. Considering that the gratings have different harmonics of periodicity: $\vec{q}_{n,m} = (m_x q_x, m_z q_z)$, different Bragg-diffraction branches appear:

$$\lambda = 2\frac{\cos(\theta_i)\, m_z/d_z - \sin(\theta_i)\, m_x/d_x}{(m_x/d_x)^2 + (m_z/d_z)^2} \tag{S.3}$$

Here $m_x = \cdots -2, -1, 0, 1, 2, \ldots$ corresponds to the diffraction order in the transverse direction, and $m_z = 1, 2, \ldots$ corresponds to the harmonics of longitudinal periodicity. The expression in Eq. S.3 bears an orientation character, showing the qualitative tendencies in the diffraction map. In order to quantitatively compare with the precise FDTD calculations (Fig. S4) and the experiments, the expression in Eq. S.3 is to be modified taking into account averaged refraction index $n$ of the filtering layers, which modifies the propagation wavevectors (and propagation angles) inside. This leads to the expression:

$$\lambda = 2\frac{\sqrt{n^2 - \sin^2(\theta_i)}\, m_z/d_z - \sin(\theta_i)\, m_x/d_x}{(m_x/d_x)^2 + (m_z/d_z)^2} \tag{S.4}$$

**Supplementary 4:** Fringing in transmission maps

The wave with modulus $|\vec{k}_0 n| = \omega/c$ diffracts in a periodic structure with reciprocal lattice vector $q_x = 2\pi m_x/d_x$ (see Fig. S5). Here, $m_x = \cdots, -2, -1, 0, 1, 2, \ldots$ corresponds to the diffraction order in the transverse direction. At some incidence angles, the diffracted mode is in resonance with the waveguiding condition of the planar structure. The resulted waveguiding mode wavevector can be defined as $k_m = 2\pi m/(2d)$. Here, $m = 1,3,5$ … corresponds to a number of the waveguiding mode and $d$ is the thickness of the waveguiding layer. Note that, due to specific excitation in a bulk of the layered structure, only odd waveguiding modes are considered.

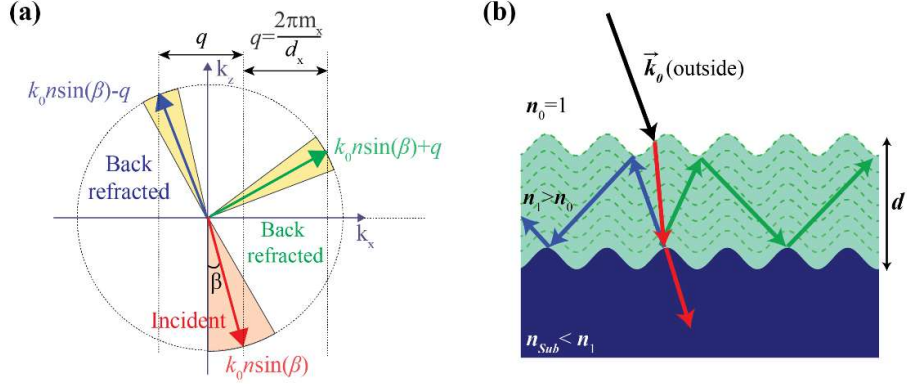

Figure S5. The schematics of (a) excitation of guided modes and (b) wavevector matching geometry for resonant excitation.

Then by matching the diffracted mode and waveguiding mode wavevector components:

$$(k_0 \sin(\theta_i) \pm q)^2 + k_m^2 = k_0^2 n^2 \tag{S.5}$$

where $\sin(\theta_i) = n\sin(\beta)$ and $n$ is the refractive index of the waveguiding layer. From the Eq. S.5 it follows:

$$k_m = \sqrt{k_0^2 n^2 - (k_0 \sin(\theta_i) \pm q)^2}. \tag{S.6}$$

Then the distance between the neighboring allowed $k_m$ modes is $\Delta k_m = k_{m+1} - k_{m-1}$ is:

$$\Delta k_m = \frac{2\pi}{d} = -\frac{(k_0 \sin(\theta_i) \pm q)k_0 \cos(\theta_i)\Delta\theta_i}{\sqrt{k_0^2 n^2 - (k_0 \sin(\theta_i) \pm q)^2}}. \tag{S.7}$$

and the angular period of the fringes follows from Eq. S.7:

$$|\Delta\theta_i| = \frac{\sqrt{k_0^2 n^2 - (k_0 \sin(\theta_i) \pm q)^2}}{(k_0 \sin(\theta_i) \pm q)k_0 \cos(\theta_i)} \frac{2\pi}{d} \tag{S.8}$$

Considering the $k_0 \sin(\theta_i) \ll q$, $\cos(\theta_i) \approx 1$ and $k_0 = 2\pi/\lambda$, Eq. S.8 can be simplified:

$$|\Delta\theta_i| = \sqrt{\left(\frac{n}{\lambda}\right)^2 - \left(\frac{m_x}{d_x}\right)^2} \frac{\lambda d_x}{m_x} \frac{1}{d}. \tag{S.9}$$

It follows from Eqs. S.8 and S.9 that the angular period of the fringes inversely proportional to the thickness of the structure. Hence, to check this interpretation, we performed numerical tests for different thickness of the layered structure (different number of layers). Parameters correspond to the fabricated structure in the main text: $n=2$, $\lambda \approx 600$ nm, $d_x=600$ nm, the number of periods: 7.5, 12.5, 16.5, 20.5 and 25.5. Figures S6(a)-S6(e) show the expected tendency and Fig. S6 (f) shows good quantitative correspondence with Eq. S.9 for the first-order diffraction $m_x=1$. Further, for the proposed structure having 33-layers and 20-periods in the $x$-direction, the electric field distribution measurements are conducted with full FDTD analyses. For the two specific wavelength values corresponding to on- and off-fringe transmission points, electric field profiles are presented in Figs. S6(g) and S6(h), respectively.

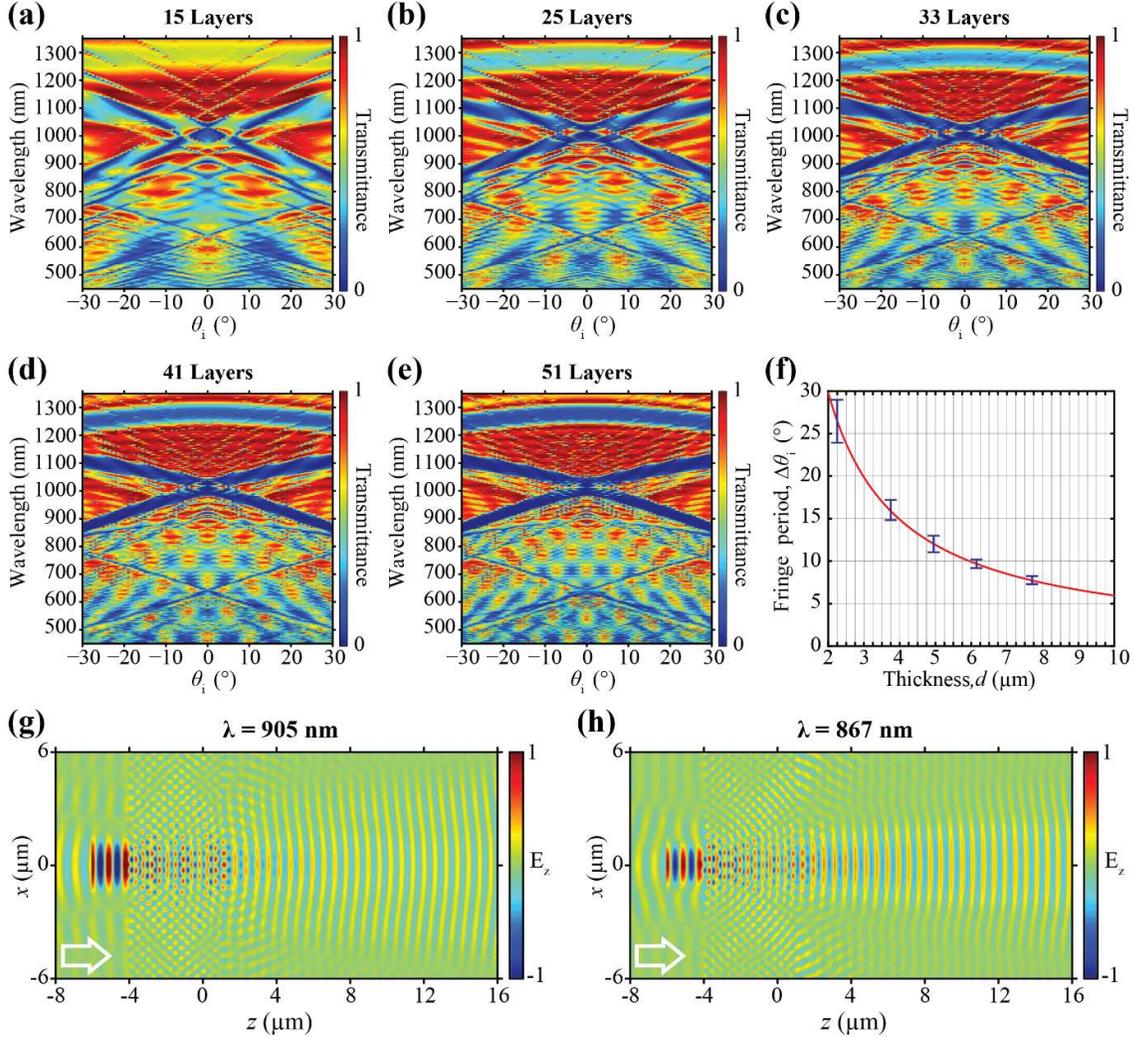

Figure S6. Zero-order transmittance maps calculated for the structure with a different number of layers (a-e). (f) Comparison of the angular fringing period with the analytic estimation Eq (S.9). (g, h) Full FDTD electric field profile calculations for the structure having 33 layers at the wavelengths corresponding to the minimum and maximum transmission (on- and off-fringe), respectively. The arrows indicate the propagation direction of the incident wave.

**Supplementary 5:** Sample preparation

Here we provide the details of the fabrication of modulated grating. The process starts with the fabrication of master grating. The basis for the master grating were commercially available lithographic plates: PHOTO-R (provided by Geola Digital UAB, Vilnius). The third harmonics of Nd:YAG laser radiation ($\lambda$ = 355 nm, pulse duration 3.1 ns) was self-interfered on a positive photoresist plate. Latter, exposed plates were developed in a diluted KOH/water solution to reveal the corrugated surface – master grating.

This polymeric grating was not suitable for direct application due to sensitivity to temperatures above 70 °C, absorption even in the visible range, etc. As a result, Ormocomp [1] was chosen due to its transparency (near 100%) at λ = 900-1000 nm and a nominal thermal stability at temperatures up to 270 °C. It is a hybrid polymer designed for UV imprinting.

To transfer the surface pattern from the master a soft lithography approach was used. A thermally cured Polydimethylsiloxane (PDMS) stamp was produced. A special procedure was used to reduce the surface modulation amplitude change during the process. First a standard 1:10 by weight mix of thermal initiator and PDMS precursor (DOWSIL Sylgard® 184 kit) was produced. The batch was left to precure slightly in a refrigerator for 24h at a temperature of 3-6 °C. This procedure increased the viscosity of the prepolymer. The precured mass was carefully poured onto the master facing upwards. Such sample was kept at 50 °C for 12-24 h to completely cure. The remaining PDMS prepolymer was further precured in a refrigerator for another 24h. A clean microscope slide was used as a backing substrate for the stamp. Such backing was needed to reduce surface bending deformations. The remaining PDMS prepolymer, which was of relatively high viscosity was poured onto the slide, microbubbles were removed inside a vacuum desiccator. The master structure with the cured PDMS layer was placed onto the slide with the PDMS material touching. Curing was performed at 50 °C overnight. The PDMS stamp had good enough adhesion to the slide for the master to be debonded mechanically using a scalpel tip.

For the copy fabrication step, the Ormocomp material was dropcast onto the stamp. The stamp was placed onto a hotplate preheated to 50 °C. This decreased the viscosity of the hybrid polymer somewhat. A clean soda-lime glass microscope slide was placed onto the stamp and the polymer was allowed to settle for a few seconds. A UV LED with a central wavelength of 365 nm, radiative power of ~4.5 W and an FWHM intensity angle 110° was placed 2-3 cm above the sample for 15 seconds. The sample was then turned upside down then cured for another 15 seconds. The stamp and the copy were mechanically debonded. The final structures were validated using an AFM with a resultant modulation height of around 230 nm.

**Supplementary 6:** Effects of modulation sweep

Here we show the effect of the modulation depth sweep through the photonic structure by numerical FDTD simulations. The conclusion following from the Fig. S7 is that the structure of Bragg filtering line is very weakly affected by that sweep and can safely be simulated by the periodic structure (equal thickness of layers) with the average period.

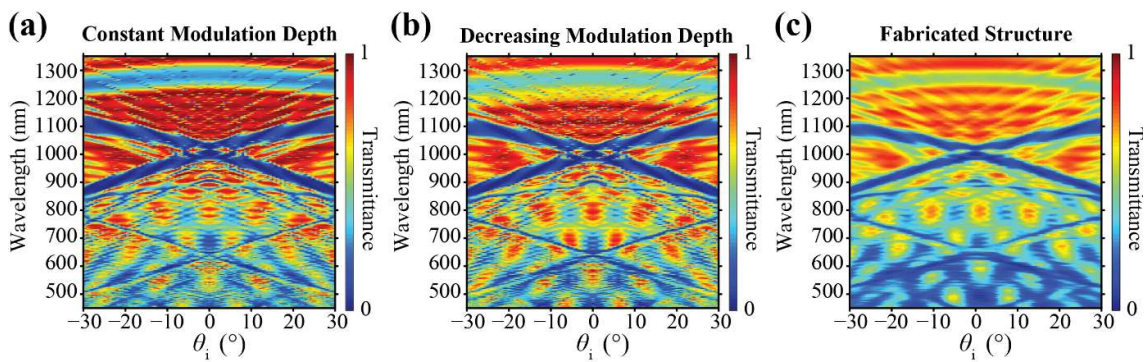

Figure S7. Zero-order transmittance maps calculated for the structure with (a) the constant amplitude of modulation (modulation depth=150 nm) (b) the modulation amplitude varying from 220 nm to 120 nm. (c) Measured zero-order transmittance map for the fabricated structure.

**Supplementary 7:** Effect of corrugation of the substrate

Comparison of numerically calculated and experimentally measured angular transmission dependences shows, that the latter are "blurred". We identify the blurring effect with the corrugation of the substrate, and subsequently deposited layered structure. The corrugation can be seen from the scanning electron microscope images of the cross-sections of the samples, as well as from the atomic force microscopy measurements of the surface of the structure, see Fig. S7. The spectrum of the surface plot Fig. S7, indicates that the background is corrugated, also that the period of the structure is corrugated as well.

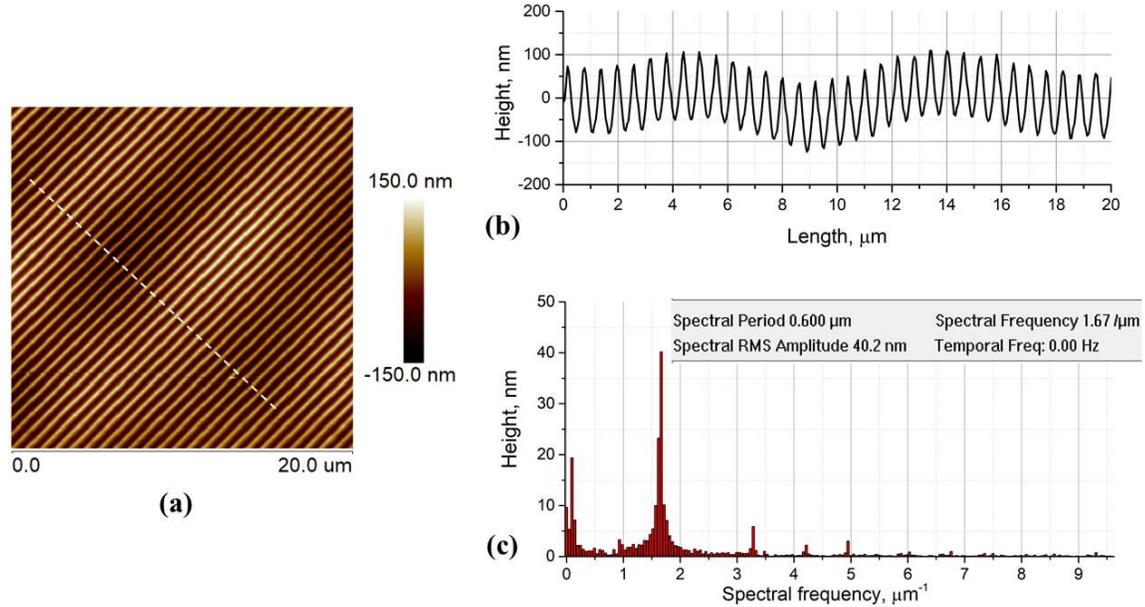

Figure S8. (a) Atomic force microscopy measurements of the surface showing the corrugation. The sample with maximal corrugation is taken, for better visibility. (b) Height distribution of the surface on a cross-section indicated (c) spectrum of the surface coordinate, showing the broadening of spectra around main harmonics (period corrugation), the spectral components around zero frequency (background corrugation) and higher harmonics (non-sinusoidal modulation).

To examine the effect in detail, we performed a numerical analysis with three different corrugation models: background corrugation, period corrugation and combination of both (see Fig. S9). Here, instead of supercell calculations, we employed full FDTD analysis by simulating a structure with 15-periods and using a Gaussian source. The transmission maps are presented in Figs. S9(b-e). Generally, the blurring effect from introduced corrugations appears in the transmission maps. Further, by taking a horizontal and vertical cross-section at a certain wavelength ($\lambda$=612 nm) and incidence angle ($\theta_i$ =2°) the effect of the corrugations is presented in Figs. S7(f) and S7(g). Especially, from Fig. S7(f), it is clearly seen that the sharpness of the transmission peak decreases with the introduced corrugations as a consequence of blurring.

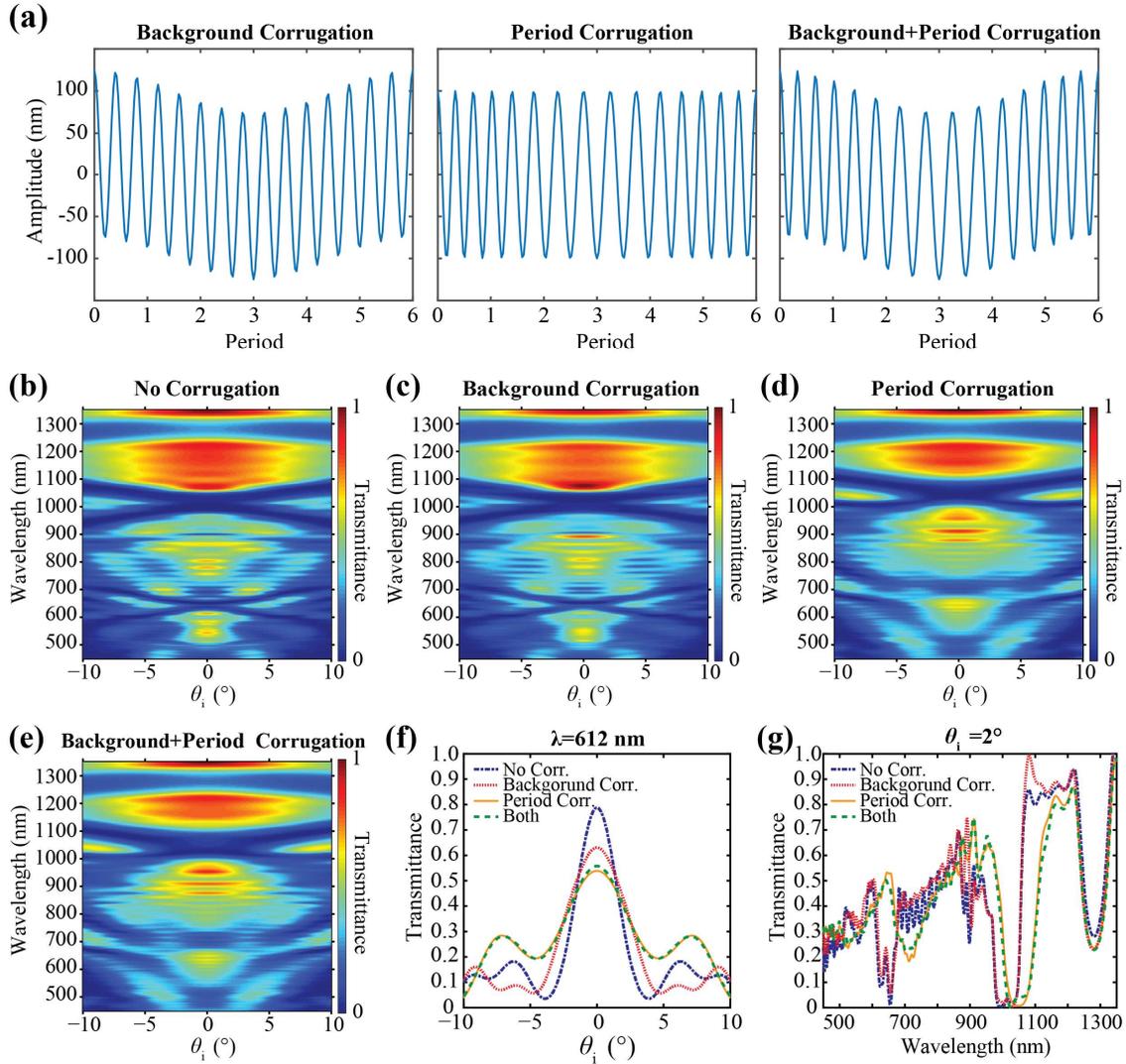

Figure S9. (a) Schematic representation of background corrugation, period corrugation and their combination, respectively. Full FDTD transmittance map calculations for the structure with (b) no corrugation, (c) background corrugation, (d) period corrugation, (e) combination of background and period corrugation. (f) superimposed horizontal cross-section profiles for all cases at the wavelength λ=612 nm. (g) superimposed vertical cross-section profiles for all cases at the incidence angle $\theta_i$=2°.

**Supplementary 8:** Beam transmission data

Here, in Fig. S10, the far-field transmitted beam intensity patterns are provided. This data was used to calculate the transmission map provided in manuscript Fig. 5. The filtering action is observed horizontally, whereas vertically the beam is affected by only some structure imperfections.

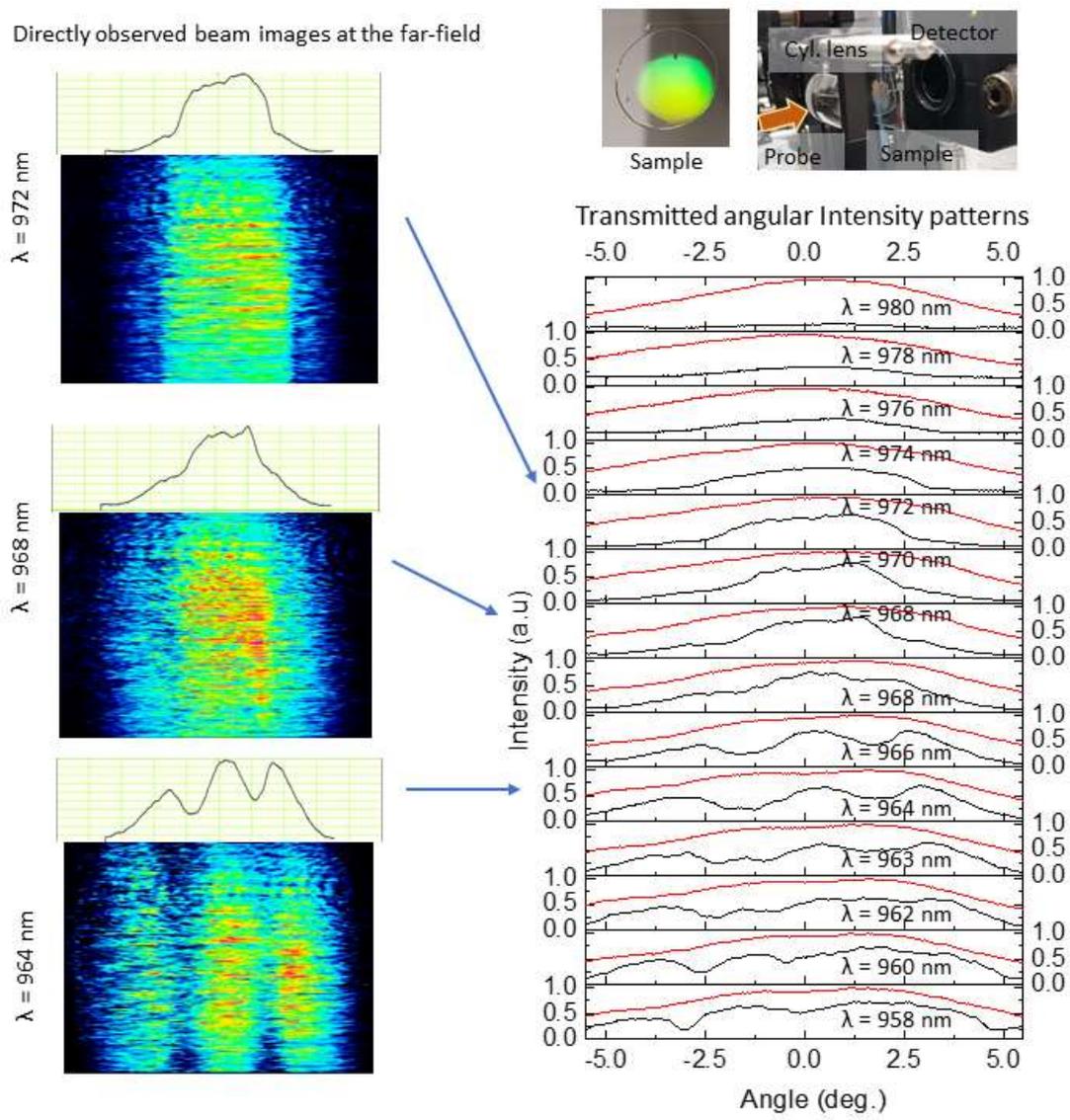

Figure S10. Detailed experimental data for spatial filtering of a divergent probe beam. Lefts images are beam profiles observed on the CMOS sensor. Right images are photo of the measurement and the angle calibrated intensity patterns averaged for approximately 300 pixels vertically with a pixel size of 4.4 micrometers.

**References**

1. "OrmoComp® and OrmoClear®FX. UV-curable Hybrid Polymers For Micro and Nano Optical Components", https://www.microresist.de/en/?jet_download=2494 , Last retrieved on 2020-08-19.